\documentclass[preprint,aps,showpacs]{revtex4}
\begin{document}
\preprint{UCI-TR 2004-24}
\title{Noncommuting spherical coordinates}
\author{Myron Bander\footnote{Electronic address: mbander@uci.edu}
}
\affiliation{
Department of Physics and Astronomy, University of California, Irvine,
California 92697-4575}

\date{July\ \ \ 2004}

\begin{abstract} Restricting the states of a charged particle to the
lowest Landau level introduces a noncommutativity between Cartesian
coordinate operators. This idea is extended to the motion of a charged
particle on a sphere in the presence of a magnetic
monopole. Restricting the dynamics to the lowest energy level results
in noncommutativity for angular variables and to a definition of a
noncommuting spherical product. The values of the commutators of
various angular variables are not arbitrary but are restricted by the
discrete magnitude of the magnetic monopole charge. An algebra,
isomorphic to angular momentum, appears. This algebra is used to
define a spherical star product. Solutions are obtained for dynamics
in the presence of additional angular dependent potentials.
\end{abstract}
\maketitle
Noncommutativity between operators corresponding to space coordinates
on a plane can be brought about via two, not totally disconnected,
procedures.  In the first case we replace the ordinary product between two
functions by the Moyal star product \cite{Moyal}
\begin{equation}\label{cartstar}
f(x)\star g(x)=\left .\exp\left(i\frac{\theta^{ab}}{2}
  \partial^{(x)}_a\partial^{(y)}_b
  \right )f(x)g(y)\right |_{y=x}\, ;\label{starproduct}
\end{equation} $\theta_{ab}$ is an anti-symmetric tensor. The second
approach consists of having a particle move on a plane in the presence
very strong, constant magnetic field perpendicular to the
plane. Letting the ratio of strength the magnetic field to the mass of
the particle approach infinity forces the system to lie in the lowest
Landau level. Restricting the dynamics to this level permits us to
treat one of the planar coordinates as a momentum conjugate to the
other one and thus introduce a noncommutativity between coordinate
variables \cite{{Bigatti:1999iz}, {Jackiw:2001dj}, {Jackiw:2002wd},
{Szabo:2004ic}}. In this work we extend this second approach to motion
of particles on a sphere, namely to noncommutativity between angular
variables. For this purpose we consider a particle of charge $e$ and
mass $\mu$ moving on a sphere of radius $r$ in the presence of a
magnetic field due to a monopole of charge $q/e$; the Dirac
quantization condition limits $q$ to the values $n/2$ where $n$ is an
integer. In the northern patch, the one excluding the south pole, the
Hamiltonian is \cite{Wu:1976ge}
\begin{equation}\label{hamiltonian0}
H=\frac{1}{2\mu r^2}\left\{p_{\theta}^2+\frac{\left[p_{\phi}
  -q(1-\cos\theta)\right]^2}{\sin^2\theta}\right\}\, .
\end{equation}

The simple approach would be to consider the above Hamiltonian in the
limit $\mu\rightarrow 0$ where we obtain the constraints $p_\theta=0$
and $p_\phi=q(1-\cos\theta)$ which in turn would imply the commutator
\begin{equation}\label{simplecomm1}
\left[\cos\theta,\phi\right]=\frac{i}{q}\, .
\end{equation}
In the Cartesian case the right hand side of the above takes on any
value inversely proportional to the strength of the applied magnetic
field. In the present situation these values are restricted by the
discrete possibilities of the magnetic monopole charge. For functions
periodic in $\phi$ this maybe rewritten as 
\begin{equation}\label{simplecomm2}
\left[\cos\theta,e^{i\phi}\right]=-\frac{e^{i\phi}}{q}\, .
\end{equation} 
Multiplying both sides by $\sin\theta$ we obtain a commutator of
variables well defined on a sphere
\begin{equation}\label{simplecomm3}
\left[\cos\theta,\sin\theta e^{i\phi}\right]=-\frac{\sin\theta
e^{i\phi}}{q}\, .
\end{equation}

We shall obtain a version of (\ref{simplecomm3}) in a more rigorous
way be considering the algebra of spherical harmonics to restricted to
the lowest level of (\ref{hamiltonian0}). Wu and Yang
\cite{Wu:1976ge,Wu:1977qk} 
studied this problem extensively and wave functions and
their properties are discussed in these references. The eigenvalues of 
(\ref{hamiltonian0}) are $E_{q;l,m}=[l(l+1)-q^2]/(2\mu r^2)$, with
$l=|q|,|q|=1,|q|+2,\ldots$ and $-l\le m\le l$; each level is $(2l+1)$ 
fold degenerate with eigenvalues being the monopole harmonics \cite{Tamm},
$Y_{q;l,m}(\theta,\phi)$. The lowest eigenvalue,
$E_{q;q,m}=q/(2\mu r^2)$, is separated by $2(q+1)/(2\mu r^2)$
from the next level. Thus in the limit $\mu\rightarrow 0$ we may
restrict the dynamics to the lowest level with states $|q;q,m\rangle$.
As most expressions depend on $|q|$ we shall treat the case $q>0$

To this end we define a spherical $q$-product
\begin{equation}\label{spqprod}
\langle q;q,m_2|(f(\theta,\phi)\cdot g(\theta,\phi))_q|q;q,m_1\rangle=\sum_m
\langle q;q,m_2|f(\theta,\phi)|q;q,m\rangle\langle
 q;q,m|g(\theta,\phi)||q;q,m_1\rangle\, ,
\end{equation}
where
\begin{equation}\label{matelem}
\langle q;q,m'|f(\theta,\phi)|q;q,m\rangle=\int
Y^*_{q:q,m'}(\theta,\phi)f(\theta,\phi)
Y_{q:q,m}(\theta,\phi)d\Omega\, .
\end{equation}
Eq.~(\ref{simplecomm3}) suggests that we look at the matrix elements
of $Y_{1,m}(\theta,\phi)$ in the level $l=q$. All such expressions may
be found in \cite{Wu:1977qk}.
\begin{equation}\label{matelem1}
\langle q;q,m_2|Y_{1,m}(\theta,\phi)|q;q,m_1\rangle=(-1)^{m_2+1-q}(2q+1)
\sqrt{\frac{3}{4\pi}}\left(
\begin{array}{ccc}
q&1&q\\-q&0&q\end{array}\right)\left(\begin{array}{ccc}
q&1&q\\-m_2&m&m_1\end{array}\right)\, ;
\end{equation}
where the arrays are Wigner $3j$ symbols and $m=m_2-m_1$. Explicit expressions
for these $3j$ symbols are readily available \cite{3j} yielding
\begin{equation}\label{matelem2}
\langle q;q,m_2|Y_{1,m}(\theta,\phi)|q;q,m_1\rangle=
\frac{(-1)^{m+1}}{q+1}\sqrt{\frac{3}{4\pi}}\left\{
\begin{array}{ccc}
\sqrt{(q+m_2)(q-m_1)}&{\rm for}&m=1\\
m_1&{\rm for}&m=0\\
-\sqrt{(q-m_2)(q+m_1)}&{\rm for}&m=-1\, .
\end{array}
\right.
\end{equation}
Using (\ref{spqprod}), the $q$-commutator is defined as 
\begin{equation}\label{starcomm1}
\left[f(\theta,\phi),g(\theta,\phi)\right]_q=(f(\theta,\phi)\cdot 
g(\theta,\phi))_q-(g(\theta,\phi)\cdot f(\theta,\phi))_q
\end{equation}
we obtain
\begin{equation}\label{starcomm2}
\left[Y_{1,0}(\theta,\phi),Y_{1,1}(\theta,\phi)\right]_q=-\frac{1}{q+1}
\sqrt{\frac{3}{4\pi}}Y_{1,1}(\theta,\phi)\, ,
\end{equation}
which agrees with (\ref{simplecomm1}) for large $q$. The
$q$-commutator of $Y_{1,1}$ with $Y_{1,-1}$ is 
\begin{equation}\label{starcomm3}
\left[Y_{1,1}(\theta,\phi),Y_{1,-1}(\theta,\phi)\right]_q=\frac{2}{q+1}
  \sqrt{\frac{3}{4\pi}}Y_{1,0}(\theta,\phi)\, .
\end{equation}
From (\ref{matelem2}) or from (\ref{starcomm2}) and (\ref{starcomm3})
we find that under the spherical $q$-product the $Y_{1,m}$'s form
an algebra isomorphic to angular momentum with
\begin{eqnarray}\label{Lcorresp1}
(q+1)\sqrt{\frac{4\pi}{3}}Y_{1,1}&\leftrightarrow& L_+\, ,\nonumber \\
(q+1)\sqrt{\frac{4\pi}{3}}Y_{1,0}&\leftrightarrow& -L_z\, ,\\
(q+1)\sqrt{\frac{4\pi}{3}}Y_{1,-1}&\leftrightarrow& -L_-\nonumber
\end{eqnarray}
or equivalently,
\begin{equation}\label{Lcorresp2}
-(q+1){\bf{\hat r}}\leftrightarrow {\bf L}\, .
\end{equation}

In addition to non trivial commutation relations for angular
position operators, we would like to obtain a definition of a star
product for these variables. Such a star product will agree with the
$q$-product only for commutators but not for simple products
\cite{Jackiw:2001dj, Szabo:2004ic}. We do require that a star product
reduce to an ordinary one when multiplying commuting variables; the
$q$-product does not do that. For Cartesian coordinates the most
direct way of obtaining a star product in (\ref{cartstar}), consistent with
$[r_a,r_b]=i\theta_{ab}$, is through the Fourier transform. Namely,
\begin{equation}\label{fourstar}
e^{i{\bf k}\cdot{\bf r}}e^{i{\bf q}\cdot{\bf r}}= e^{\frac{i}{2}\theta^{ab}k_aq_b}
e^{i({\bf k+q})\cdot{\bf r}}\, .
\end{equation}
For the angular case, we must modify the product of two spherical
harmonics to allow for noncommuting angular variables. To this end we
start with an unconventional expression for the coefficient of $Y_{L,M}$
in the expansion of the product of two spherical harmonic (usually
written as a product of to $3j$ symbols), $\int Y_{l_1,m_1}({\bf{\hat
r}})Y_{l_2,m_2}({\bf{\hat r}}) Y_{L,M}^*({\bf{\hat r}})d{\bf{\hat
r}}$. From the expansion of a plane wave in terms of spherical waves
we find
\begin{equation}\label{FourBes}
Y_{l,m}({\bf{\hat r}})=\frac{i^{-l}}{4\pi j_l(kr)}\int 
  e^{i{\bf k}\cdot{\bf x}}Y_{l,m}({\bf{\hat k}})d{\bf{\hat r}}\, ;
\end{equation}
this expression is independent of the magnitudes of ${\bf k}$ and
${\bf r}$.
The previously discussed expansion coefficient becomes
\begin{equation}\label{expcoef1}
\int Y_{l_1,m_1}({\bf{\hat r}})Y_{l_2,m_2}({\bf{\hat r}})
Y_{L,M}^*({\bf{\hat r}})d{\bf{\hat r}}=\frac{i^{-l_1-l_2}}{j_{l_1}(kr)j_{l_2}(qr)}
\int e^{i{\bf k}\cdot{\bf r}} e^{i{\bf q}\cdot{\bf r}}
  Y_{l_1,m_1}({\bf{\hat k}})Y_{l_2,m_2}({\bf{\hat
  q})}Y_{L,M}^*({\bf{\hat r}})d{\bf{\hat r}}d{\bf{\hat k}}d{\bf{\hat
  q}}\, .
\end{equation}
When the components of ${\bf{\hat r}}$ commute with each other the
product of the two exponentials in the above integrals is treated
normally. In the noncommuting situation we have to define such a product
to be consistent with the commutators in (\ref{starcomm2}) and (\ref{starcomm3}).
Using the correspondence in (\ref{Lcorresp2}) we may make the replacement
\begin{equation}\label{Lrrepl}
\exp(i{\bf k}\cdot{\bf r})\rightarrow \exp\left(-ir
  \frac{1}{q+1}{\bf k}\cdot{\bf L}\right)
\end{equation} 
with a similar expression for $\exp(i{\bf q}\cdot{\bf r})$. The
product of the two exponentials is treated as a product of two
rotations. The result is then inserted into (\ref{expcoef1}) to obtain
the desired coefficients. This time the result depends on the
magnitudes $k$ and $r$ indicating that, as in the Cartesian case,
different star products will result in the same star commutator.

Following Peierls \cite{Peierls}, who studied the problem of a charged
particle that, in addition to the strong magnetic field, is acted on
by some potential, we can add an angle dependent potential, $V(\theta,\phi)$ to the
present problem. In general, the solution requires the diagonalization
od a $(2q+1)\times (2q+1)$ matrix. In the simple case 
$V(\theta,\phi)=\lambda\cos\theta$ the eigenstates are still the
$|q;q,m\rangle$'s  and the corresponding energies are
\begin{equation}\label{cosener}
E_{q;q,m}=q/(2\mu r^2)-(-1)^m\frac{\lambda m}{q+1}\, .
\end{equation}

\end{document}